\documentclass[12pt, peerreview , onecolumn,letterpaper]{IEEEtran}
\pdfoutput=1

\RequirePackage[normalem]{ulem} 
\RequirePackage{color}\definecolor{RED}{rgb}{1,0,0}\definecolor{BLUE}{rgb}{0,0,1} 

\RequirePackage{color}

\usepackage{tikz}
\usepackage[sumlimits,intlimits]{amsmath}
\usepackage{amsmath,amsfonts,amssymb}%
\usepackage{bm}%
\usepackage{graphicx,graphics}%
\usepackage{cases}%
\usepackage[noadjust]{cite}%
\usepackage{color}%
\usepackage{cite,url}
\usepackage{verbatim}
\usepackage{enumerate}
\usepackage{algorithm}
\usepackage{balance}
\usepackage{multirow}
\usepackage{stfloats}
\usepackage{subfigure}
\usepackage{xspace}
\usepackage{wrapfig}
\usepackage{smartref}

\newcommand{\inches}{\ensuremath{{}^{\prime\prime}}}

\ifCLASSOPTIONpeerreview

\fi

\begin{document}

\title{Improving Bandwidth Efficiency in E-band \vspace{-16pt}\\ Communication Systems}


\author{\vspace*{-8pt}
\authorblockN{Hani Mehrpouyan,~\IEEEmembership{Member, IEEE},
                M. Reza Khanzadi,~\IEEEmembership{Student Member, IEEE,}
                Michail Matthaiou,~\IEEEmembership{Member, IEEE},
                Akbar M. Sayeed,~\IEEEmembership{Fellow, IEEE,}
                Robert Schober,~\IEEEmembership{Fellow, IEEE,}
                and
                Yingbo Hua,~\IEEEmembership{Fellow, IEEE}
              }\\
}
\maketitle
\thispagestyle{empty}
{\let\thefootnote\relax\footnotetext{{Hani Mehrpouyan is with the Department of Electrical and Computer Engineering, California State University. M. Reza Khanzadi and Michail Matthaiou are with the Department of Signals and Systems, Chalmers University of Technology, Sweden. Akbar M. Sayeed is with the Department of Electrical and Computer Engineering at the University of Wisconsin, Madison. Robert Schober is with the Department of Electrical and Computer Engineering at University of British Columbia, Canada. Yingbo Hua is with the Department of Electrical Engineering at the University of California, Riverside. Emails: hani.mehr@ieee.org, khanzadi@chalmers.se, michail.matthaiou@chalmers.se, akbar@engr.wisc.edu, rschober@ece.ubc.ca, and yhua@ee.ucr.edu.
\vspace{-0pt}}} } 

\vspace*{-0pt}
\begin{abstract}

\vspace*{-0pt}

The allocation of a large amount of bandwidth by regulating bodies in the 70/80 GHz band, i.e., the \emph{E-band}, has opened up new potentials and challenges for providing affordable and reliable Gigabit per second wireless point-to-point links. This article first reviews the available bandwidth and licensing regulations in the \emph{E-band}. Subsequently, different propagation models, e.g., the ITU-R and Cane models, are compared against measurement results and it is concluded that to meet specific availability requirements, E-band wireless systems may need to be designed with larger fade margins compared to microwave systems. A similar comparison is carried out between measurements and models for oscillator phase noise. It is confirmed that phase noise characteristics, that are neglected by the models used for narrowband systems, need to be taken into account for the wideband systems deployed in the E-band. Next, a new multi-input multi-output (MIMO) transceiver design, termed continuous aperture phased (CAP)-MIMO, is presented. Simulations show that CAP-MIMO enables E-band systems to achieve fiber-optic like throughputs. Finally, it is argued that full-duplex relaying can be used to greatly enhance the coverage of E-band systems without sacrificing throughput, thus, facilitating their application in establishing the backhaul of heterogeneous networks.


\end{abstract}

\newpage

\vspace*{-30pt}
\section{Introduction}

The popularity of multimedia applications and broadband internet has created an ever increasing demand for achieving higher throughputs in cellular and wireless networks. Thus far, wireless point-to-point links have been playing an important role in carrying a large portion of this data by interconnecting cellular base stations or enterprise buildings. In fact, due to their low cost of installation and insusceptibility to environmental effects, more than fifty percent of today's cellular base stations are interconnected using wireless backhaul links \cite{Huang-12-M}. Yet, if wireless point-to-point links are expected to continue to be widely applied in next generation cellular networks, they have to support throughputs comparable to that of fiber-optic links. This task is made difficult by the limited bandwidth available in the microwave band \cite{Wells-12-M}. In this regard, the large bandwidth available in the $70$ and $80$ GHz or \emph{E-band} has opened up new opportunities for developing multi-{Gigabit per second (Gbps)} wireless links \cite{Pi-12-M,Huang-12-M}.


Even though the available bandwidth in the E-band is more than fifty times the entire cellular spectrum, radio signals in the E-band are more adversely affected by environmental factors \cite{wells_10}. The characteristics of E-band signals and systems can be summarized as follows:
\begin{itemize}
\item Due to the higher carrier frequencies, the antennas are more directional, making E-band systems mainly suitable for \emph{line-of-sight (LOS)} applications.\vspace{+3pt}
\item Rain and obstacles more severely attenuate radio signals in the E-band. Consequently, with the same transmit power and link availability requirements, E-band wireless links can operate over shorter distances when compared to microwave systems. For example, let us consider two point-to-point wireless systems with a $99.999\%$ availability requirement and a fade margin of $0$ dB: the first system operating at $23$ GHz and employing $256$-quadrature amplitude modulation (QAM) can achieve a link distance of $3$ kms at $1.4$ Gbps, while the second system utilizing the $70$/$80$ GHz spectrum and using binary phase shift keying (BPSK) can only operate over $1.9$ kms at $3$ Gbps \cite{wells_10}. \vspace{+3pt}

\item To achieve the high carrier frequencies required by E-band systems, a voltage controlled oscillator's signal needs to be taken to E-band carrier frequencies using a larger frequency multiplication factor compared to systems operating in the microwave band. This, in turn, can result in larger oscillator phase noise variances. Phase noise, which is present in communication systems due to imperfect oscillators, can significantly impact their bandwidth efficiency and performance, since it results in the rotation of the signal constellation from one symbol to the next symbol \cite{article_PHASE_N_MODEL_I}. Moreover, in E-band systems, due to the LOS nature of the links, the coherence time of the channel is much longer compared to the phase noise variation time. This means that phase noise can be a performance bottleneck in E-band systems whilst in other systems the channel variations might be the fundamental limitation.\vspace{+3pt}

\item Because of the received signal's large bandwidth and high sampling rate, E-band systems require the application of high speed digital signal processing, digital-to-analog conversion (D/A), and analog-to-digital conversion (A/D) units at the transceivers. \vspace{+3pt}

\item Due to the very high carrier frequencies, the power amplifiers used in E-band systems have a very limited output range and are inefficient compared to those employed in the microwave band. Hence, the output power levels of most existing E-band systems are lower than the maximum levels allowed by regulating bodies. This further limits the operating range of these systems.
\end{itemize}
Because of these limitations, thus far, most E-band systems use low order modulations such as BPSK and on-off keying, and are not spectrally efficient compared to traditional microwave links. In fact, current E-band systems achieve a spectral efficiency of $0.5$--$2.4$ b/s/Hz \cite{article_Dyadyuk_09}, whereas the spectral efficiencies of traditional microwave systems are in the range of $4$--$12$ b/s/Hz \cite{receiver_eband-2}.

To enable the development of multi-Gbps wireless links, it is paramount to introduce new transceiver designs for E-band systems that can more efficiently utilize the available bandwidth, while supporting wireless links over distances comparable to those of microwave links. To this end, this article first reviews the bandwidth allocation and licensing in the E-band. Next, unlike previous articles that did not take into account the effect of phase noise \cite{article_Csurgai_12,article_Dyadyuk_09}, by comparing measurement results with the current models for signal attenuation and oscillator phase noise, it is shown that traditional models developed for the microwave band may not accurately predict these phenomena in the E-band. The development of more accurate models is anticipated to result in better link budget planning and more accurate tracking of phase noise, which can in turn enhance the bandwidth efficiency of E-band systems, e.g., enabling the application of higher order modulations. Subsequently, a new multi-input multi-output (MIMO) transceiver design, termed \emph{continuous aperture phased MIMO (CAP-MIMO)}, is outlined and new topologies and applications for E-band systems are proposed that can mitigate their limitations and better utilize their potential.

The remainder of this article is organized as follows: In Section \ref{sec_potential}, the available bandwidth and the regulations in the E-band are summarized. Section \ref{sec:bw_eff} focuses on the models for the channel attenuation and oscillator phase noise for the E-band. In Section \ref{sec:eband-MIMO}, the CAP-MIMO transceiver design for E-band systems is presented while Section \ref{sec:app} proposes new topologies and applications for E-band systems.


\vspace*{-12pt}

\section{Bandwidth and Licensing in the E-band}\label{sec_potential}
The International Telecommunication Union (ITU) has allocated the $71$--$76$ and $81$--$86$ GHz bands for establishing wireless links throughout the world. Nevertheless, the channelization and regulations regarding the use of this spectrum differ in different regions and countries.



\vspace*{-12pt}
\subsection{United States}

In the United States, the entire $10$ GHz of bandwidth in the $71$--$76$ and $81$--$86$ GHz bands is available for utilization without any specific sub-channelization. This approach provides a great potential for deployment of very high speed frequency division duplex (FDD) wireless links. In addition, the Federal Communications Commission (FCC) set the effective isotropic radiated power (EIRP), the maximum transmit power, the minimum antenna gain, and the maximum out of band emissions for E-band systems to $55$ dBW, $35$ dBm, $43$ dBi, and $-13$ dBm, respectively. More importantly, due to the very high carrier frequencies and the requirement of high antenna directivity, wireless links in the E-band can be deployed in close vicinity of one another with limited interference. Therefore, the FCC has adopted a unique licensing approach for spectrum allocation in the E-band, where the links can be quickly and economically registered over the Internet.

\vspace*{-12pt}
\subsection{Europe}
In Europe, the Conference of European Postal and Telecommunications has taken a different approach and has divided the $10$ GHz of spectrum in the $70$/$80$ GHz bands into $250$ MHz channels with a guard band of $125$ MHz at the top and bottom end of each band. Thus, each $5$ GHz band consists of $19$ channels that can be used for both FDD and time division duplex operation. Moreover, the regulations in the E-band in Europe are more stringent than the FCC rules. Specifically, in Europe the EIRP, the maximum transmit power, the minimum antenna gain, and the maximum out of band emissions are set to $55$ dBW, $30$ dBm, $38$ dBi, and $-30$ dBm, respectively, where the EIRP and the maximum transmit power are also functions of the antenna gain and are expected to be lowered as the antenna gain increases.

\vspace*{-12pt}
\subsection{Asia}
The $70$/$80$ GHz bands are under consideration for licensing in Japan, China, and most other Asian countries and are not yet available for commercial use.


\vspace*{-2pt}
\section{Channel and Phase Noise Models for the E-band}\label{sec:bw_eff}

Accurate channel and oscillator phase noise models for the E-band spectrum are essential for link budget
\begin{wrapfigure}{r}{0.55\textwidth}
\begin{center}
\vspace*{-20pt}
\scalebox{0.44}{\includegraphics {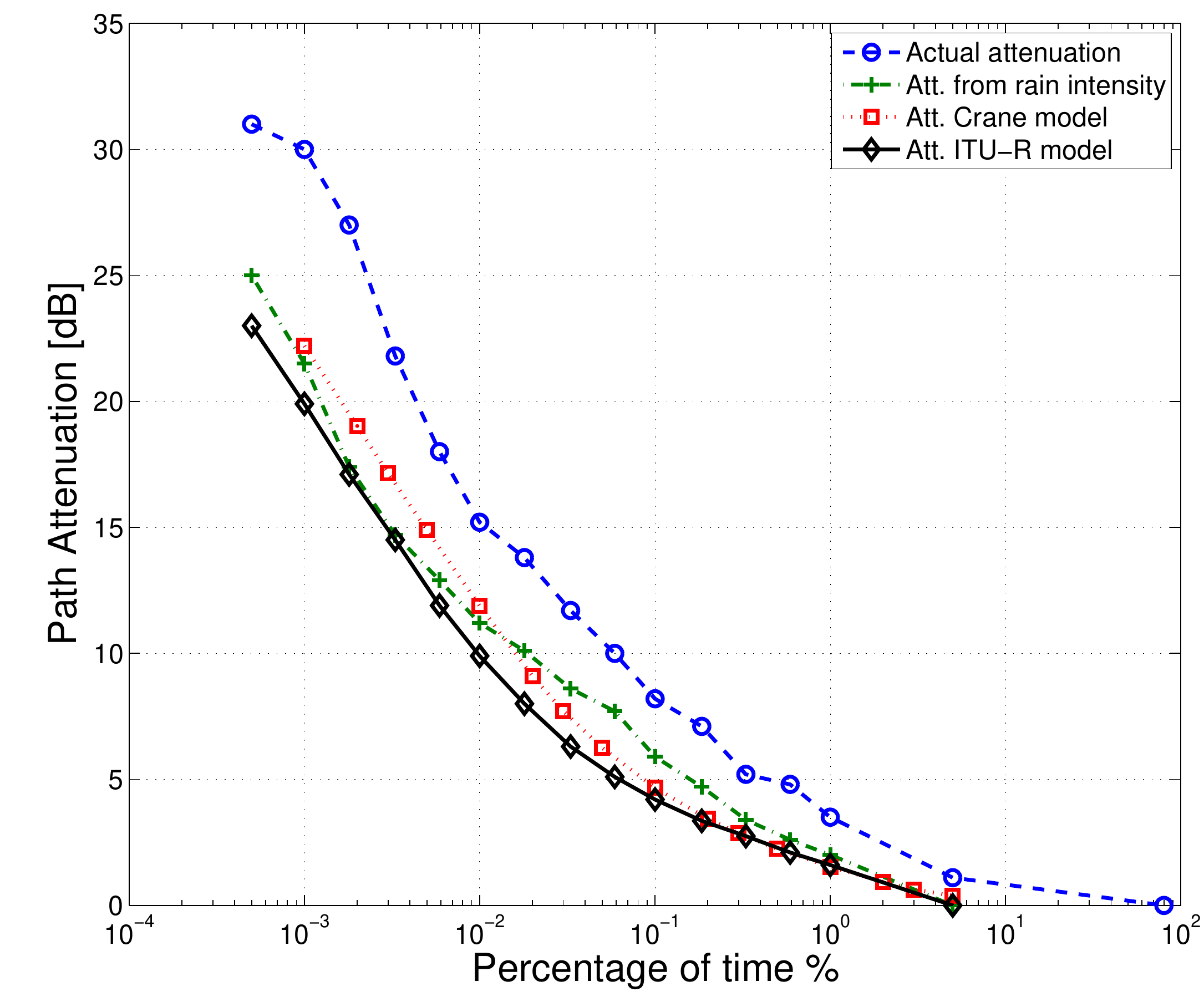}}
\vspace{-20pt}
\centering
\caption{Path attenuation versus time with transmit power$=$$18.6$ dBm, receiver threshold$=$$-58$ dBm, antenna gain$=$$43$ dBi, Tx/Rx separation$=$$1$ km $\pm50$ m, data rate$=$$1.25$ Gbps, differential BPSK (the measurements are also reported in \cite{Hansryd-10}).
}
\label{fig:los_MIMO}
\vspace{-30pt}
\end{center}
\end{wrapfigure}
planning and accurate tracking of phase noise in E-band systems. Both of these improvements are expected to enhance the bandwidth efficiency of E-band systems in the near future \cite{receiver_eband-2}. Thus, in this section, we examine the accuracy of the existing models for both phenomena in the E-band.

\vspace*{-10pt}
\subsection{Channel and Propagation Characteristics in E-band}\label{sec:chnl_prop}

To accurately predict the effect of environmental conditions on the performance of wireless communication systems, the ITU-R and Crane models have been extensively applied for link-budget planning in the microwave band. In order to determine the accuracy of these models in predicting rain intensity and the resulting signal attenuation in the E-band, a long term measurement campaign was carried out in Gothenburg, Sweden, by Ericsson Research, where the rain intensity and the signal attenuation of an E-band system was measured over a period of nine months \cite{Hansryd-10}. The results of this measurement campaign and a comparison with respect to both the ITU-R and Crane models are plotted in Fig.~\ref{fig:los_MIMO}. It can be observed that both the ITU-R and Crane models can rather sufficiently predict the rain intensity in this region, since the signal attenuation calculated based on the measured rain intensity is close to that of the ITU-R and Crane models. However, the measurement results in Fig.~\ref{fig:los_MIMO} show that the measured signal attenuation in the E-band is considerably higher than the attenuation predicted by both the ITU-R and Crane models. This demonstrates that both models, which are well suited for the microwave band, are not capable of accurately predicting the channel attenuation in the E-band. Thus, for E-band wireless point-to-point links to meet the expected $99.999\%$ availability requirements, the fade margin needs to be chosen larger than the values calculated using the Crane and ITU-R models, e.g., $5$--$10$ dB higher as shown in Fig.~\ref{fig:los_MIMO}. This new finding indicates that to avoid larger than necessary fade margins, more accurate channel attenuation models have to be developed for the E-band. These more accurate channel and propagation models are also anticipated to enhance the bandwidth efficiency of E-band systems.

Note that although the shortcomings of the ITU-R model in predicting the attenuation for E-band systems has also been confirmed in \cite{article_Csurgai_12}, in this work, for the first time, we present a comparison with respect to the Crane attenuation model, which is more extensively applied in North America.\footnote{The work in \cite{article_Csurgai_12} also provides a more accurate model compared to the ITU-R model for characterizing the channel and propagation characteristics of wireless communication systems in the E-band.}

\vspace*{-10pt}
\subsection{Phase Noise Models in E-band}\label{sec:eband-pn}

One of the main challenges in E-band communication systems is to equip the transceivers with low phase noise high-frequency oscillators. Nevertheless, recent studies have shown that such oscillators may be designed using Gallium-Nitride technology, waveguide theory, and opto-electronic techniques \cite{Russer1998,Ruscito2011}. Generally, two methods are proposed to generate high-frequency oscillation \cite{Russer1998,Ruscito2011}. One is to design an on-chip high-frequency oscillator and the other is to increase the frequency of a low-frequency oscillator by means of frequency multipliers. Even though the former is expected to result in more accurate oscillators, research has shown that the design of accurate and affordable oscillators for commercial applications via this approach is a challenging task. On the other hand, the latter approach increases a low-frequency oscillator's phase noise variance by the so-called multiplication factor \cite{Russer1998}. As a result, oscillator phase noise is one of the main limiting factors in the application of higher order modulations in E-band systems \cite{wells_10}.

Although the effect of oscillator phase noise in narrowband systems has been extensively studied, there is a lack of understanding of this impairment for the wideband systems deployed in the E-band. According to traditional phase noise models, the phase noise variance or rate is linearly proportional to the sampling time applied at the receiver \cite{article_PHASE_N_MODEL_I}. Therefore, it is expected that systems employing larger bandwidths and smaller sampling times will experience a smaller phase noise variance. However, by increasing the signal bandwidth, other system parameters such as the bandwidth of the receiver front-end filter must also be increased. Consequently, this leads to an increase in phase perturbation introduced to the entire communication system.
\begin{figure}[!h]
    \centering
       \begin{minipage}[h]{0.46\textwidth}
            \centering
            \vspace{-0.1in}
            \hspace{-.8 cm}
            \includegraphics[width=1.05\textwidth]{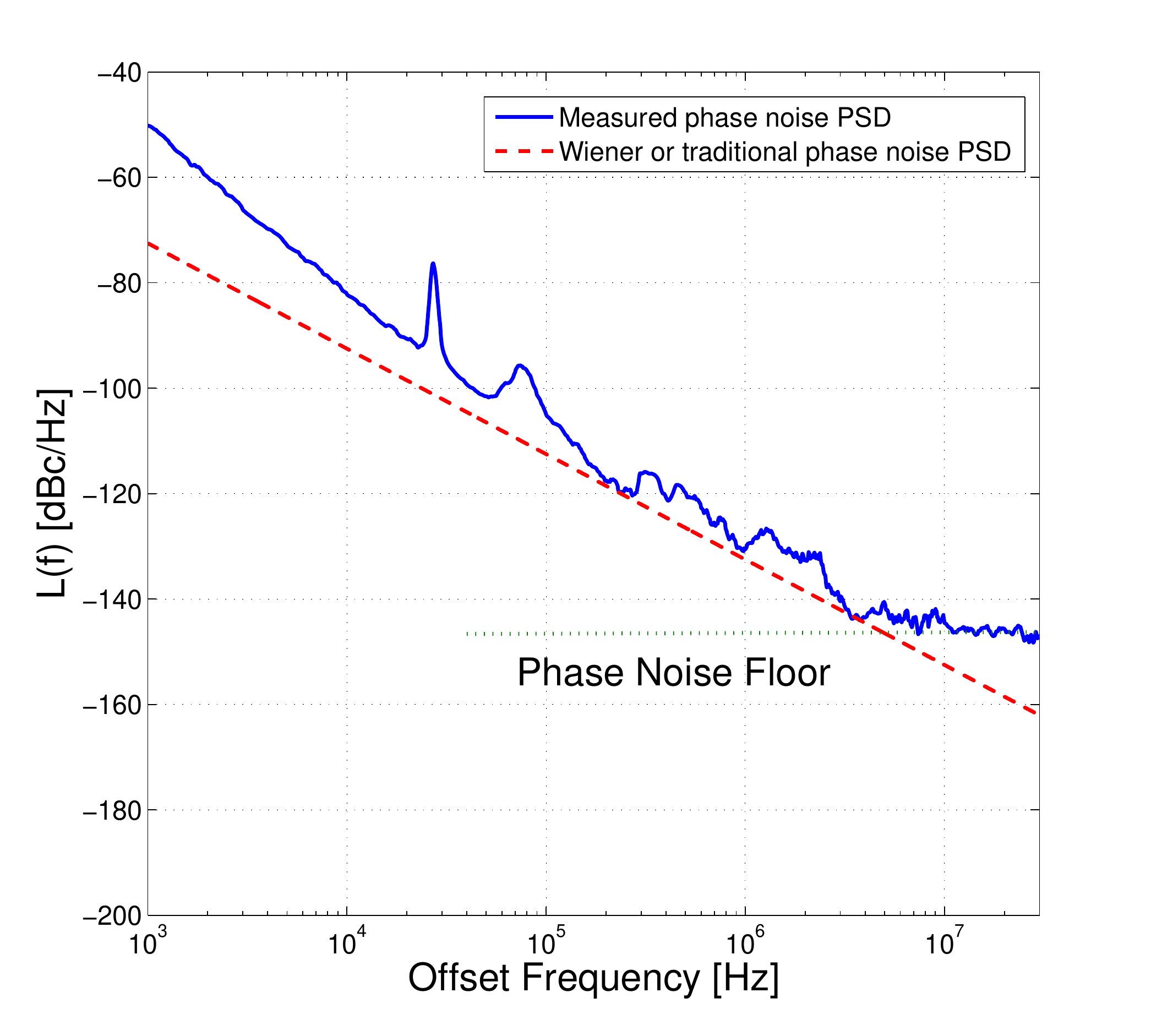}
            \vspace{-.29 in}
            \caption{The power spectral density of a free-running \newline oscillator operating at 9.9 GHz.}
            \label{fig:OSC_PSD}
        \end{minipage}
        \begin{minipage}[h]{0.46\textwidth}
            \centering
            \includegraphics[width=0.98 \textwidth]{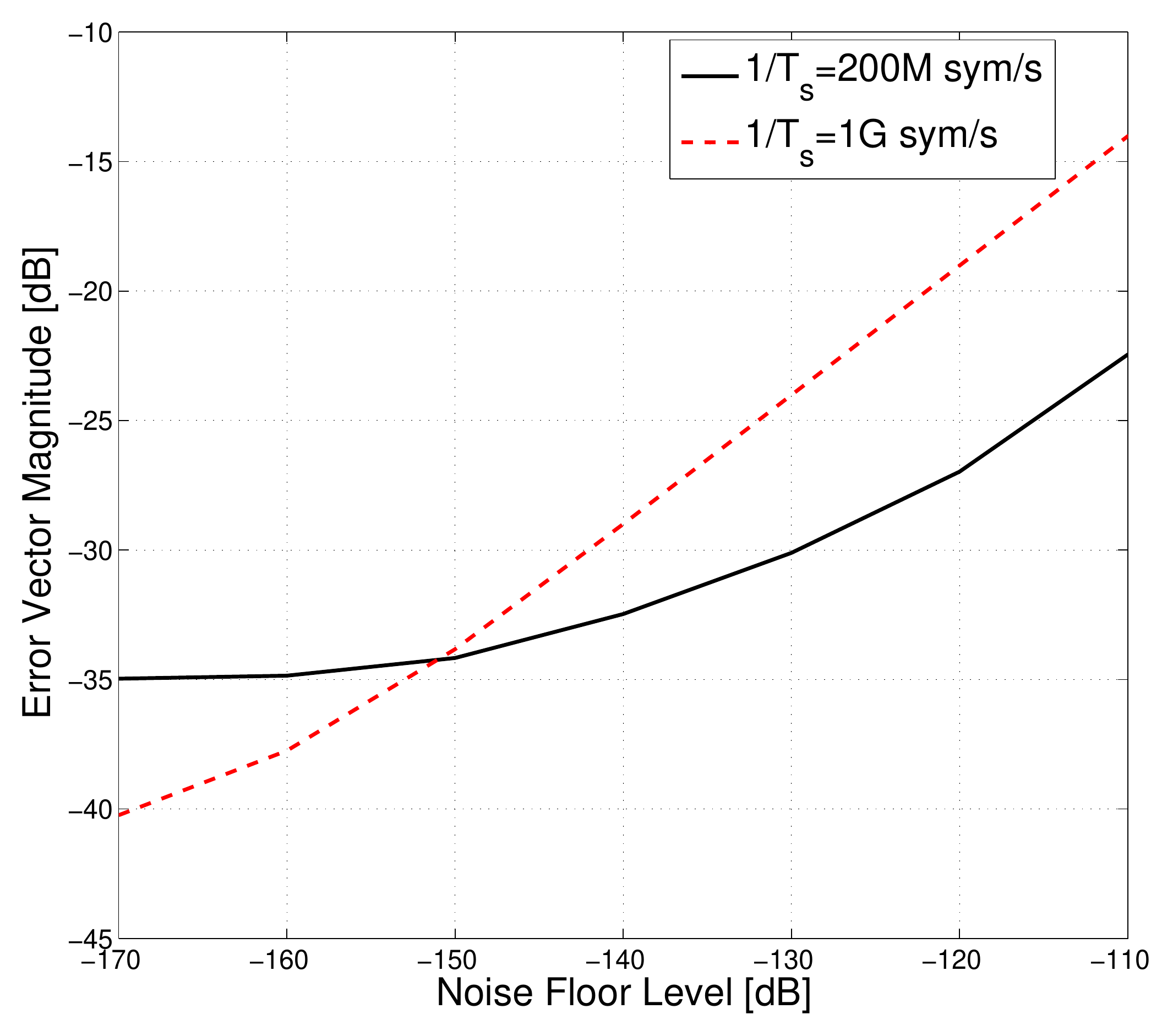}
            \vspace{-.20 in}
            \caption{EVM of residual phase noise error for two systems with different symbol rates vs. oscillator phase noise floor power.}
            \label{fig:GSMvsCDMA2}
        \end{minipage}
  \vspace{-0.0 in}
\end{figure}

To illustrate this effect in wideband systems, Fig.~\ref{fig:OSC_PSD} depicts the measured power spectral density (PSD) of oscillator phase noise for a monolithic microwave integrated circuit oscillator operating at $9.9$ GHz. Fig.~\ref{fig:OSC_PSD} compares measurement results with the traditional Wiener phase noise model, which is extensively applied for the microwave band. Note that unlike the traditional phase noise models, the phase noise PSD of an oscillator does not continue to decrease with increasing offset frequency. In fact, as shown in Fig.~\ref{fig:OSC_PSD}, in practice, the phase noise PSD exhibits a floor region beyond a certain offset frequency.\footnote{This effect is not exclusive to this oscillator and is also reported in \cite{article_PHASE_N_MODEL_I} and references therein.} Thus, as the bandwidth of a communication system increases, the floor region in the PSD of an oscillator is expected to play an important role in the overall system's performance. To verify this finding, in Fig.~\ref{fig:GSMvsCDMA2}, the performances of two communication systems with different bandwidths are compared in terms of the \emph{error vector magnitude (EVM)} (also known as receive constellation error) for different phase noise floor levels (signal-to-noise ratio (SNR)$=30$ dB and $16$-QAM). Fig.~\ref{fig:GSMvsCDMA2} shows that below a certain noise floor level, the cumulative phase noise is dominant and a system with higher symbol rate experiences a lower EVM. However, as the phase noise floor increases, the performance of a communication system employing a larger bandwidth degrades more dramatically. It is also important to consider that due to the use of frequency multipliers, it is expected that most oscillators used in E-band systems have higher phase noise floor levels. For example, if the above $9.9$ GHz oscillator is used in an E-band system operating at $70$ GHz, a frequency multiplier with a multiplication factor of $7$ needs to be applied, which will increase the noise floor by $20\log7=16.9$ dBc. Therefore, from the results in Figs. \ref{fig:OSC_PSD} and \ref{fig:GSMvsCDMA2}, it can be concluded that the phase noise models that are used for narrowband systems, e.g., the Wiener model, cannot be applied to accurately predict the properties of this impairment in wideband systems. Thus, more accurate phase noise models for wideband communication systems have to be developed to estimate and compensate the effect of phase noise more effectively, which in turn enables the use of more bandwidth-efficient modulation schemes.

%


\vspace{-4pt}
\section{MIMO Transceiver Design for E-band Systems}\label{sec:eband-MIMO}

The development of MIMO technology has been largely based on the assumption of rich multipath which combined with the deployment of multiple antennas results in multiple independent spatial channels between two terminals. Under these circumstances, it has been theoretically shown that the MIMO system capacity scales linearly with the minimum of the number of transmit and receive antennas. However, E-band systems are expected to operate under strong LOS conditions, thereby creating several research challenges and opportunities for the design of efficient E-band MIMO transceivers \cite{Sayeed-10,Pi-12-M,article_MIMO_LoS}. More importantly, the antenna properties in the E-band are attractive for three important reasons:
\begin{enumerate}
\item they result in high antenna gains for a given antenna size,
\item they enable highly directional communication with narrow beams, thus, reducing interference, and
\item they support the deployment of large-dimensional MIMO systems with relatively compact antenna arrays.
\end{enumerate}


Two benchmark E-band systems dominate the current state-of-the-art. In the first configuration, termed a \emph{DISH} system, conventional continuous aperture ``dish" antennas are used in highly directional LOS point-to-point links. Such systems are currently used for wireless backhaul links, e.g., in the commercial systems offered by Siklu, E-band Communications, or Bridgewage. In the second configuration, termed \emph{conventional MIMO}, the antenna elements are placed sufficiently far apart so that the spatial LOS responses become independent. The required antenna spacing can be worked out via simple geometrical arguments, and leads to the so-called Rayleigh spacing. For a given transmitter-receiver distance, the Rayleigh antenna spacing is inversely proportional to the carrier frequency. Thus, compared to microwave systems, LOS MIMO technology is more suitable for E-band systems, since the antenna spacing is smaller and the transceivers can be housed within a relatively compact module. However, while such systems can exploit multiplexing gains, they suffer from poor power efficiency and increased interference \cite{Sayeed-10}. In principle, the above limitations of conventional MIMO systems can be eliminated by using half-wavelength spaced large antenna arrays. Although such systems can optimally exploit the spatial dimension, they suffer from a prohibitively high transceiver complexity due to the requirement of a large number of array elements. For example, a $6\inches$ planar array operating at $80$ GHz requires about $6400$ antenna elements, while each antenna element requires a dedicated transceiver module \cite{Sayeed-10}.

Recall that propagation in the E-band is expected to have sparse multipath components and is predominantly LOS. Thus, the the spatial multiplexing gain of a MIMO E-band system can, in practice, be much smaller than the minimum of the number of transmit and receive antennas employed. In other words, compared to a system in the microwave band, the spatial communication subspace for a MIMO E-band system can be expressed with a smaller number of orthogonal basis functions \cite{Sayeed-10,Pi-12-M}. Accordingly, to fully exploit the potential of MIMO technology and to reduce the transceiver complexity, the number of beams transmitted or received by a MIMO E-band system needs to be equal to the dimensionality of the E-band channel subspace. This characteristic of E-band channels has motivated the development of the \emph{CAP-MIMO} transceiver.

CAP-MIMO combines the multiplexing gain of MIMO systems, the antenna gain of DISH systems, and the beamforming capability of phased arrays to optimally exploit the smaller spatial dimensionality at E-band frequencies \cite{Sayeed-10}. CAP-MIMO uses a
\begin{figure}[t]
\begin{center}
\vspace*{-5pt}
\scalebox{0.60}{\includegraphics {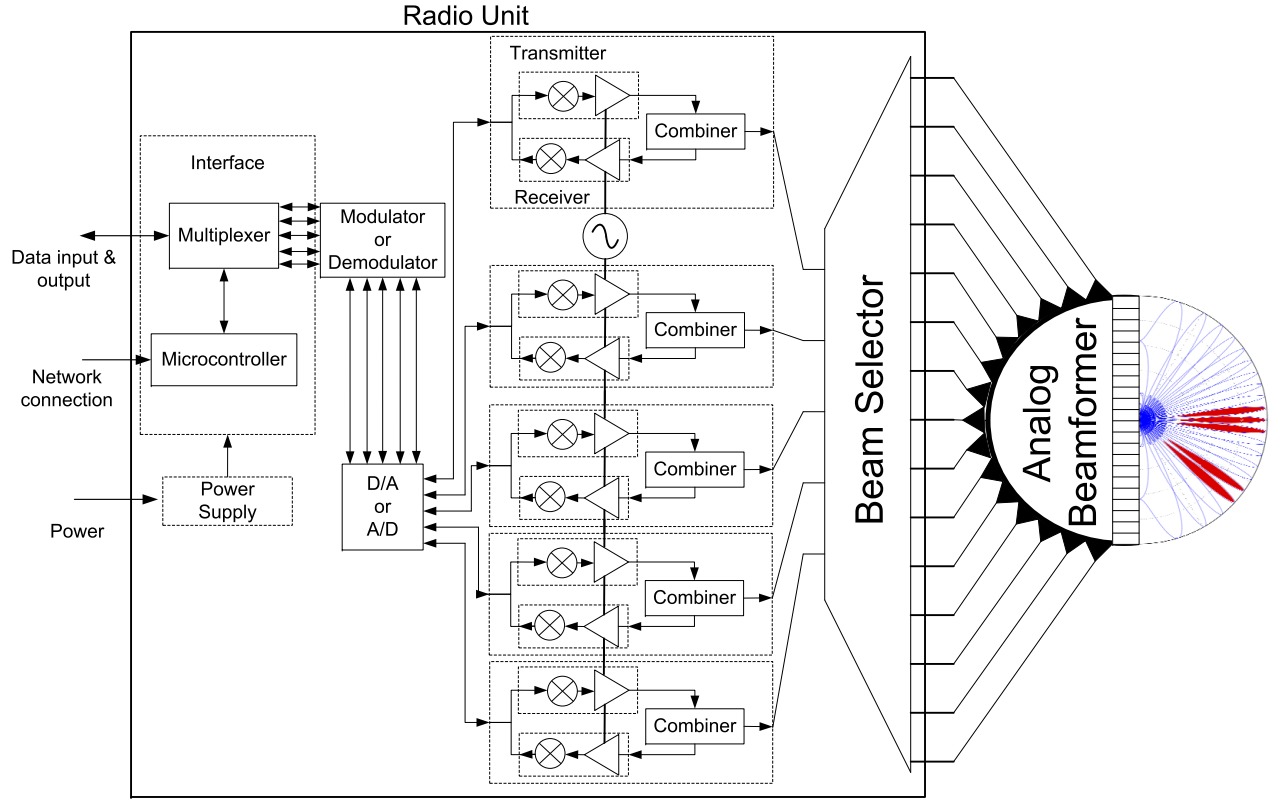}}
\vspace{-80pt}
\caption{Radio unit for an E-band CAP-MIMO system.}
\vspace*{-10pt}
\label{fig:radio}
\end{center}
\end{figure}
high-resolution discrete lens array to perform analog beamforming in the passband, see Fig.~\ref{fig:radio}. Essentially, in this setup, a relatively small number of active beams are radiated by the corresponding feed antennas on the focal surface of the lens array. The number of transmitted beams is directly proportional to the dimensionality of the communication channel. This approach ensures that the CAP-MIMO transceiver is equipped with the smallest number of A/D, D/A, and radio frequency units, while fully taking advantage of the potential of MIMO technology. Fig.~\ref{fig:radio} depicts the radio unit for a CAP-MIMO system. In this setup, it is assumed that $5$ spatially independent channels can be established between transmitter and receiver. Consequently, the CAP-MIMO system only requires $5$ transceiver blocks. The beam selector block in Fig.~\ref{fig:radio} ensures that appropriate beams are selected for signal transmission and reception, which is analogous to an antenna selection block in a conventional MIMO system.





\begin{wrapfigure}{r}{0.60\textwidth}
\begin{center}
\vspace*{-20pt}
\hspace*{-5pt}\scalebox{0.39}{\includegraphics {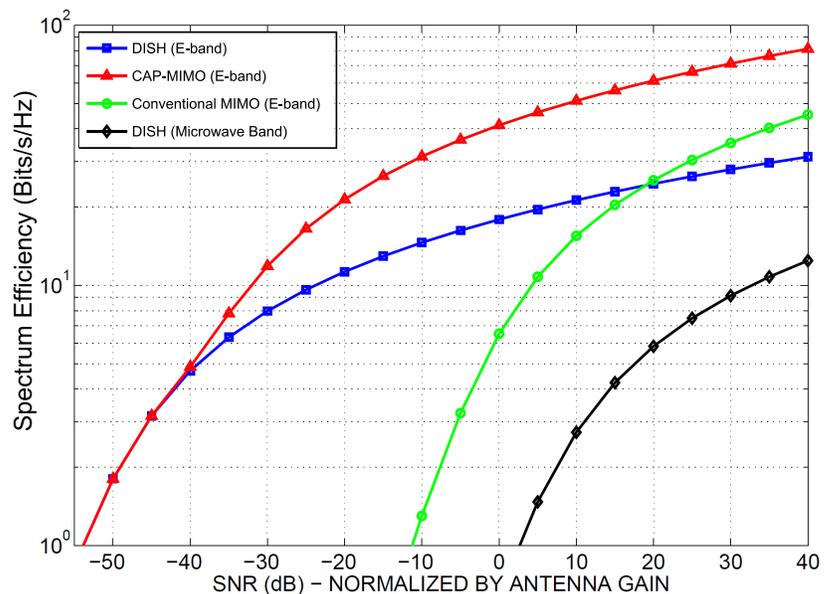}}
\vspace{-30pt}
\centering
\caption{Bandwidth efficiency comparison of CAP-MIMO, conventional MIMO, and DISH systems.}
\label{fig:capacity}
\vspace{-20pt}
\end{center}
\end{wrapfigure}

Fig.~\ref{fig:capacity} compares the bandwidth efficiency of CAP-MIMO with a conventional MIMO system with widely spaced antennas, and a DISH system for a point-to-point LOS link at $80$ GHz. The link length is about $200$ m. The antenna size for the DISH and CAP-MIMO systems is $0.6$ m $\times$ $0.6$ m with an antenna gain of $55$ dBi. The conventional MIMO system has $4$ widely spaced antennas at each end,
where each antenna has a gain of $30$ dBi. The CAP-MIMO system is assumed to take advantage of $4$ transmit and receive beams. One important observation is that the DISH system is optimum at low SNRs, whereas conventional MIMO outperforms DISH at high SNRs. CAP-MIMO yields the best performance across the entire SNR range, thereby representing a robust scheme for realizing the advantages offered in the E-band. CAP-MIMO can achieve a spectral efficiency of $30$ bits/s/Hz, at a normalized SNR of $-8$ dB with an SNR gain of about $25$ dB, over the other two systems. This results in a data rate of $30$--$300$ Gbps for a bandwidth of $1$--$10$ GHz. For the given antenna size, a system operating in the $3$ GHz band, i.e., the microwave band, at best achieves a spectral efficiency of $10$ bits/s/Hz at a much higher SNR of $40$ dB. This corresponds to a maximum data rate of $5$ Gbps when considering a generous bandwidth of $500$ MHz. Finally, CAP-MIMO also offers a promising route to electronic multi-beamforming and steering which can be exploited for a number of attractive operational functions, such as user tracking in mobile environments, multi-beam steering in point-to-multipoint links, and formation of high capacity cooperative MIMO links (see Section \ref{sec:eband-coop}).

It is important to note that although the use of multi-antenna arrays in E-band systems has been studied \cite{article_Dyadyuk_09}, they are only used for power combining and not MIMO spatial multiplexing gain. As such, the CAP-MIMO concept presented here promises to provide spectral efficiencies that are significantly higher than the approach in \cite{article_Dyadyuk_09}, e.g., $30$ bits/s/Hz for CAP-MIMO compared to $4.8$ bits/s/Hz for the approach in \cite{article_Dyadyuk_09}. Moreover, this article uniquely provides a comprehensive comparison between the throughput of traditional MIMO and DISH systems at different SNRs.

\vspace{-10pt}
\section{E-band Point-to-Point Systems in Next Generation Cellular Networks}\label{sec:app}

With respect to point-to-point wireless links, e.g., backhaul links, the most important issues for cellular providers are range, very high link availability, and operating costs. To facilitate the application of E-band systems in next generation cellular networks, we propose to employ full-duplex relaying to extend the range of E-band links. Moreover, the challenges and potentials for applying E-band systems for backhauling in heterogeneous networks (HetNets) are presented.

\vspace{-10pt}
\subsection{Full-Duplex Multi-Hop Cooperative Relay Networks}\label{sec:eband-coop}

\begin{figure}[t]
\begin{center}
\vspace*{-5pt}
\scalebox{0.60}{\includegraphics {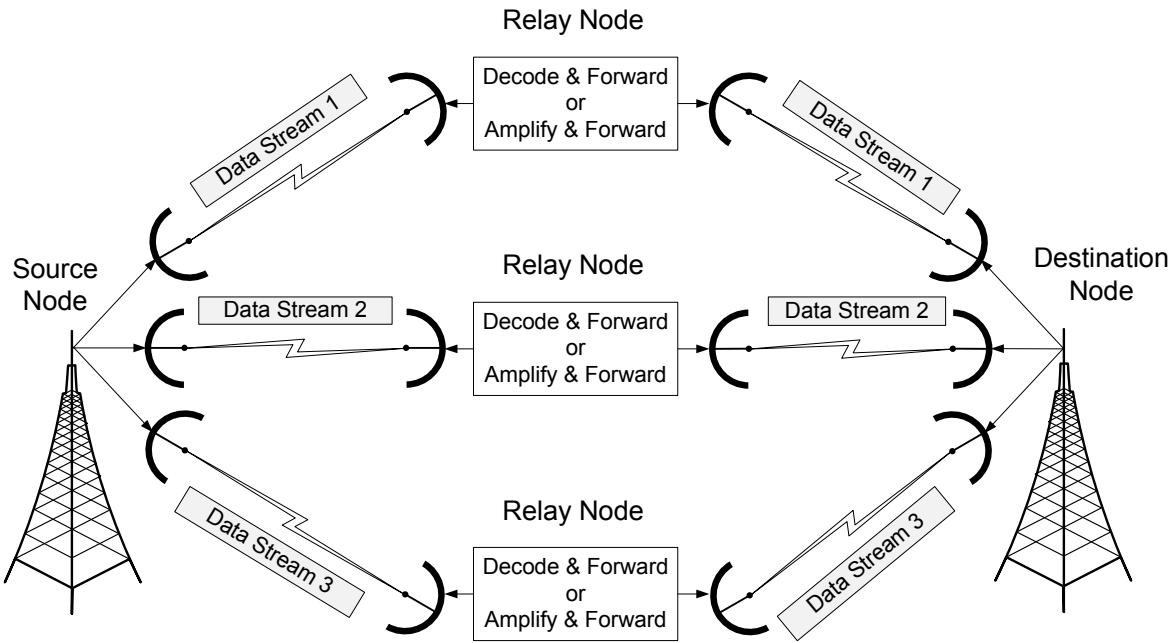}}
\vspace{-120pt}
\caption{A full-duplex cooperative relay network in the E-band.}
\vspace*{-10pt}
\label{fig:relay-network}
\end{center}
\end{figure}

The research in the field of cooperative relay networks has shown that such systems can significantly enhance the coverage area and reliability of point-to-point wireless links \cite{Sheng-11-A}. Regenerative and non-regenerative relaying, i.e., decode-and-forward (DF) and amplify-and-forward (AF) relaying, respectively, have gained significant traction due to their performance and simplicity. However, most of the research in this area has been focused on half-duplex cooperative relaying, where one node transmits while the remaining nodes stay silent. This approach reduces the throughput of the network for every additional hop. Moreover, although full-duplex relaying circumvents the shortcomings of half-duplex relaying, its implementation in the microwave band has been a challenging task. One approach is for the relays to transmit and receive at difference frequency bands. However, such an approach is bandwidth inefficient and halves the potential system capacity for each additional hop. Another approach is to ensure that each node transmits and receives simultaneously over the same bandwidth. Nevertheless, this scheme can result in significant interference at the transceiver, which degrades system performance.


Recall that E-band systems employ highly directional antennas. Thus, unlike traditional microwave systems, E-band transceiver modules can operate simultaneously without significant interference. Moreover, due to the LOS nature of E-band links and the higher atmospheric absorption, E-band wireless devices can be deployed in close vicinity of one another while maintaining a low level of interference, i.e., higher spatial reuse factor. These are essential features that can enable the development of full-duplex multi-relay cooperative networks in the E-band. Without sacrificing throughput, such networks can greatly enhance the range of E-band point-to-point links, see Fig.~\ref{fig:relay-network}. Moreover, AF relaying is especially attractive for full-duplex networks, since the relays may not be required to convert the signal into the baseband. Thus, significantly simplifying the relays' structures and reducing their hardware costs. For example, in the scenario in Fig.~\ref{fig:relay-network}, it is shown that by applying appropriate beamforming schemes at the transmitter and the relays, the throughput of an E-band point-to-point system can be tripled while the link distance can be doubled.


\vspace{-10pt}
\subsection{E-band Systems in Heterogeneous Networks}\label{sec:eband-hetnet}

To meet the surge in the number of users and each user's throughput requirements, cellular providers have resorted to increasing the number and density of macrocell base stations (BSs). However, as these throughput demands continue to grow, large BSs cannot effectively meet users' needs in different settings, e.g., in indoor environments and at the cell edges \cite{article_Ghosh12}. Moreover, macrocell BSs are expensive to deploy and maintain. Thus, next generation cellular networks are expected to adopt the HetNet paradigm, where smaller and more specialized cells are deployed by the operators, e.g., picocells in urban areas and femtocells in indoor environments \cite{article_Ghosh12}. Furthermore, to mitigate interference, achieve smooth hand-offs from tier to tier, and enable cooperation amongst different tiers, fast backhaul links are of paramount importance.

Since picocells and femtocells have smaller coverage areas, they can be deployed more densely throughout a cellular network. Moreover, E-band systems are capable of establishing multi-Gbps wireless links over short distances. Thus, they are well suited for establishing backhaul links amongst different tiers within a cell. Due to the high atmospheric absorption and the use of highly directional antennas, they can do so without causing significant interference.

There are, however, some challenges that need to be addressed. Most femtocells are expected to be deployed in indoor environments. Since E-band radio signals do not penetrate obstacles and buildings very well, E-band backhaul links may require the application of strategically positioned relays throughout the network to establish reliable connectivity between femtocell and macrocell BSs. Another challenging issue, is the high development cost of E-band transceivers. This is especially important for the development of next generation HetNets, since femtocell and picocell BSs are expected to be relatively inexpensive to manufacture. Nevertheless, recent research in the field of semiconductor design and fabrication, e.g., Silicon-Germanium and Gallium-Nitride, is expected to reduce the manufacturing costs of E-band systems and facilitate their adoption \cite{wells_10}.

\vspace*{-16pt}
\section{Summary}\label{sec:conc}

In this article, to gain a better understanding of the effects of channel attenuation and phase noise on E-band systems, measurement results for both phenomena were compared against existing models. It was concluded that the ITU-R and Cane models may not be able to accurately determine the fade margin for E-band systems, e.g., if these models are used to obtain the required fade margin for an E-band link with $99.999\%$ availability, the result may be $5$--$10$ dB below the actual required value. These measurements also indicated that the floor in the PSD of oscillator phase noise may more significantly influence the performance of wideband systems operating in the E-band compared to the narrowband systems in the microwave band. Further research in this field is anticipated to result in E-band transceivers that can employ higher order modulations and improved bandwidth efficiency. Next, we showed that the CAP-MIMO transceiver design can enable E-band systems to achieve bandwidth efficiencies of up to $30$ b/s/Hz (data rates of up to $300$ Gbps). Finally, it was proposed that due to application of highly directional antennas and the LOS nature of E-band links, full-duplex relaying can be used to extend the range of E-band systems and to facilitate their application in next generation heterogeneous cellular networks.




\bibliographystyle{IEEEtran}
\bibliography{IEEEabrv,Bay_PN}

\end{document}